\documentclass[reprint, 
 superscriptaddress,
 amsmath,
 amssymb,
 aps, 
 showkeys, 
 ]{revtex4-2}

\usepackage{graphicx}
\usepackage{dcolumn}
\usepackage{bm}

\usepackage[hidelinks]{hyperref}

\hypersetup{
    pdftitle={Statistical theory of electronic degrees of freedom in wave packet molecular dynamics},
    pdfauthor={Daniel Plummer, Pontus Svensson, Wiktor Jasniak, Patrick Hollebon, Sam M. Vinko, Gianluca Gregori}
}

\begin{document}

\title{Statistical theory of electronic degrees of freedom in wave packet molecular dynamics}

\author{Daniel Plummer}
\email{daniel.plummer@physics.ox.ac.uk}
\affiliation{Department of Physics, University of Oxford, Oxford OX1 3PU, United Kingdom}

\author{Pontus Svensson}
\affiliation{Department of Physics, University of Oxford, Oxford OX1 3PU, United Kingdom}
\affiliation{Institute of Radiation Physics, Helmholtz-Zentrum Dresden-Rossendorf, D-01328 Dresden, Germany}

\author{Wiktor Jasniak}
\affiliation{Department of Physics, University of Oxford, Oxford OX1 3PU, United Kingdom}

\author{Patrick Hollebon}
\affiliation{AWE, Aldermaston, Reading, Berkshire RG7 4PR, United Kingdom}

\author{Sam M. Vinko}
\affiliation{Department of Physics, University of Oxford, Oxford OX1 3PU, United Kingdom}

\author{Gianluca Gregori}
\affiliation{Department of Physics, University of Oxford, Oxford OX1 3PU, United Kingdom}

\date{August 23, 2026}

\begin{abstract}
 We derive statistical distributions for the degrees of freedom in wave packet molecular dynamics models. Specifically, a theory is developed for the width distributions of Gaussian wavepackets in both isotropic and anisotropic formulations. The resulting distribution functions show good agreement with molecular dynamics data under warm dense matter conditions, providing practical guidance for constraining the confining potential, an empirical parameter in the model. We also discuss how these distributions influence the resulting effective Coulomb interactions.
\end{abstract}

\maketitle

\section{Introduction}

The modeling of many-body quantum systems from first principles presents a formidable challenge. A particularly demanding regime is so-called warm dense matter~\cite{Vorberger2026}, which has attracted significant interest because it occurs in planetary interiors, brown dwarfs, and white dwarf envelopes \cite{Saumon2022,Didier1992,Guillot1999}, as well as during the compression phase of inertial confinement fusion capsules \cite{Hu2018}. While exact wavefunction propagation is computationally unfeasible for many-particle systems, finite-temperature quantum-statistical approaches such as path integral Monte Carlo (PIMC) can provide a range of static equilibrium properties \cite{Dornheim2023,bonitz2024pub, Militzer2021}. Alternatively, density functional theory (DFT)-based methods can be employed, which provide access to many static and dynamic properties~\cite{Moldabekov2024,Militzer2021, Kononov2022, Hu2018, bonitz2024pub,Vorberger2026}. Additionally, simple pair potentials may be used in the semi-classical regime~\cite{Filinov2004,Glosli2008,Demyanov2025}.

Wave packet molecular dynamics (WPMD) is an alternative approach based on restricting the electronic wave function to yield a computationally feasible method \cite{Feldmeier2000,Grabowski2014,Klakow1994a,Jakob2007,Lavrinenko2021}. This scheme enables simulations involving many thousands of particles, significantly reducing finite-size effects compared to PIMC- and DFT-based methods. Such an approach could also be complementary to the more established methods for a variety of additional reasons. In equilibrium, WPMD captures the joint evolution of electrons and ions over long timescales, enabling direct study of inter-species coupling beyond the Born-Oppenheimer approximation. In addition to dynamic electron properties \cite{Svensson2024pub,Morozov2009}, this allows electron-ion collisional transport to be directly modeled~\cite{Svensson2025}. Furthermore, WPMD has been employed in an ongoing debate regarding the effect of electronic fluctuations on both diffusive and acoustic ion modes in dense plasma \cite{Davis2020, Yao2021,Angermeier2021}. The method has also been extended to include bound-state wavefunctions~\cite{Ebeling1997,Plummer2026}, which allowed the latter reference to extract the self-consistent ionization states using the scheme in Ref.~\cite{Plummer2025}. In the nonequilibrium context, the method was used to study the relaxation of a two-temperature electron-proton plasma and yielded slow relaxation rates when compared with semi-classical models~\cite{Ma2019}. 

However, care must be taken to ensure that the restricted functional form of the electronic wavefunction does not lead to unphysical behavior. A primary example is the confining potential, which controls the electron extent and hence regulates many properties of the model~\cite{Plummer2026, Morozov2009, Ebeling2006}. The purpose of this work is to highlight how the statistical distribution of the additional electron degrees of freedom in WPMD may be predicted from the Hamiltonian structure of the underlying equations. This provides a practical means to assess the effect of the confining potential. Here, we focus on the statistical properties of WPMD using a Gaussian ansatz, which has been widely applied to the study of warm dense matter and partially ionized plasmas.

\section{Gaussian wavepacket models}
\label{sec:Gaussian_wavepacket_model}

We employ an anisotropic Gaussian wavepacket ansatz developed in Ref.~\cite{Svensson2023}, and subsequently used in Refs.~\cite{Svensson2024pub,Plummer2026}. The key results are briefly recalled for the purposes of this work. We start with \(N\) classical protons and \(N\) quantum electrons, and employ the mixed quantum-classical approach of Ehrenfest dynamics~\cite{Hutter2009}. For the protons, the resulting evolution is dictated by the Ehrenfest Hamiltonian,
\begin{equation} \label{eq:Ehrenfest_classical_hamiltonian} \mathcal{H} = \sum_I \frac{\boldsymbol{P}_I^2}{2M_I} + \frac{e^2}{4\pi\epsilon_0}\sum_{I<J}\frac{1}{|\boldsymbol{R}_I - \boldsymbol{R}_J|} + \mathcal{H}^{\text{eff}}. \end{equation}
{Here \(M_I\), \(\boldsymbol{R}_I\) and \(\boldsymbol{P}_I\) denote the mass, position and momentum of the \(I^{\text{th}}\) ion respectively; additionally \(e\) is the electron charge and \(\epsilon_0\) is the permittivity of free space.} The corresponding Hamiltonian equations of motion are
\begin{equation} \frac{d \boldsymbol{R}_I}{dt} = \frac{\boldsymbol{P}_I}{M_I} \quad ; \quad \frac{d \boldsymbol{P}_I}{dt} = -\frac{\partial \mathcal{H}}{\partial \boldsymbol{R}_I}. \end{equation}
 The effective Hamiltonian representing the quantum subsystem is the expectation value of the electron Hamiltonian with respect to the many-body wavefunction \(|\Psi\rangle\),
\begin{equation} \mathcal{H}^{\text{eff}} = \langle \Psi | \hat{H}_e(\{\boldsymbol{R}_I\}) |\Psi\rangle, \end{equation}
where \(\hat{H}_e\) is the electronic Hamiltonian operator. This operator depends parametrically on the ionic coordinates \(\{\boldsymbol{R}_I\}\), and includes the electron kinetic energy together with all electron interaction terms. In Ehrenfest dynamics, \(|\Psi\rangle\) is allowed to evolve in the full electronic Hilbert space. Meanwhile, the ions evolve under forces obtained from the gradient of the expectation value of the electronic Hamiltonian with respect to the ionic coordinates. This corresponds to a mean-field treatment, in which the nuclei experience the averaged influence of the electronic state rather than evolving on a single potential energy surface. In WPMD, \(|\Psi\rangle\) is parameterized by a finite set of continuous parameters \(\{q_\mu\}\) which define a manifold in the full Hilbert space. To distinguish this case we perform the replacement
\begin{equation} |\Psi\rangle \rightarrow |Q(\{q_\mu\})\rangle \quad ; \quad \mathcal{H}^{\text{eff}} \rightarrow \mathcal{H}^{\text{WP}} = \langle Q | \hat{H}_e(\{\boldsymbol{R}_I\}) |Q\rangle, \end{equation} where \(|Q\rangle\) is the parameterized wavefunction with the explicit parameter dependence suppressed and \(\mathcal{H}^{\text{WP}}\) is the corresponding wavepacket Hamiltonian. The two restricted Gaussian wavefunctions studied in this work shall now be specified.

\textbf{Anisotropic form:} A Hartree product of single-particle Gaussian orbitals is chosen, each with the functional form
\begin{equation} \label{eq:ani_gauss_wavefunction} \langle \boldsymbol{x} |q_{i}\rangle = \mathcal{N} \exp\left[-(\boldsymbol{x}-\boldsymbol{r}_{i})^T \text{M}_i (\boldsymbol{x}-\boldsymbol{r}_{i}) + \frac{ i}{\hbar} \boldsymbol{p}_{i}^T(\boldsymbol{x}-\boldsymbol{r}_{i})\right] \end{equation}
where \(\boldsymbol{r}_i, \boldsymbol{p}_i\) correspond to the expected position and momenta, respectively. Additionally, \(\mathcal{N}\) is a normalization factor \cite{Svensson2023} and \(\hbar\) is the reduced Planck's constant. The symmetric complex three-by-three shape matrix
\begin{equation} \text{M}_i = \frac{1}{4}\Sigma_i^{-1} - \frac{i}{\hbar}\Pi_i \end{equation}
is commonly decomposed into a real covariance part, \(\Sigma_i\), which is positive definite with units of length squared, and a conjugate shape-momentum variable \(\Pi_i\). Explicitly the electronic subsystem now depends on the real parameters \(\{\boldsymbol{r}_i, \boldsymbol{p}_i, \Sigma_i, \Pi_i\}_{i=1}^N\) where \(i\) labels an electron in the system. The equations of motion are of Hamiltonian form
\begin{equation} \label{eq:com_eom} \frac{d \boldsymbol{r}_i}{dt} = \frac{\partial \mathcal{H}^{\text{WP}}}{\partial \boldsymbol{p}_i} \quad ; \quad \frac{d \boldsymbol{p}_i}{dt} = -\frac{\partial \mathcal{H}^{\text{WP}}}{\partial \boldsymbol{r}_i} \end{equation}
for the vector degrees of freedom, while the matrix degrees of freedom have the following symmetrized Hamiltonian structure:
 \begin{equation} \label{eq:internal_eom}
\begin{split}
\frac{d}{dt} \Sigma_{i\beta\alpha} =& \frac{1}{2}\left(\frac{\partial \mathcal{H}^{\text{WP}}}{\partial \Pi_{i\alpha\beta}} + \frac{\partial \mathcal{H}^{\text{WP}}}{\partial \Pi_{i\beta\alpha}}\right)\\
\frac{d}{dt} \Pi_{i\alpha\beta} =& -\frac{1}{2}\left(\frac{\partial \mathcal{H}^{\text{WP}}}{\partial \Sigma_{i\alpha\beta}} + \frac{\partial \mathcal{H}^{\text{WP}}}{\partial \Sigma_{i\beta\alpha}}\right),
\end{split}
\end{equation}
where \(\Sigma_{i\alpha\beta} = [\Sigma_{i}]_{\alpha\beta}\) is the $\alpha, \beta$ component of the symmetric width matrix and, similarly, \(\Pi_{i\alpha\beta} = [\Pi_{i}]_{\alpha\beta}\) is a component of the conjugate width-momentum matrix, both corresponding to the \(i\)-th electron. 

\textbf{Isotropic form:} The isotropic case is also studied. Constraining the matrices to \begin{equation} \label{eq:ani_to_iso} \Sigma_i = s_i \text{I}\quad ; \quad \Pi_i = \pi_{s,i} \text{I}, \end{equation}
where \(\text{I}\) is the identity matrix and \(s_i, \pi_{s,i}\) are real scalars, gives the isotropic state
\begin{equation} \label{eq:iso_gauss_wavefunction} \langle \boldsymbol{x} |g_{i}\rangle = \mathcal{N} \exp\left[-m_i(\boldsymbol{x}-\boldsymbol{r}_{i})^2 + \frac{i}{\hbar} \boldsymbol{p}_{i}^T (\boldsymbol{x}-\boldsymbol{r}_{i})\right], \end{equation}
with the complex width variable
\begin{equation} m_i = \frac{1}{4s_i}-\frac{i}{\hbar} \pi_{s,i}. \end{equation} While the equations of motion for the expected position \(\boldsymbol{r}_i\) and expected momentum \(\boldsymbol{p}_i\) remain with the same form given in Eq.~\eqref{eq:com_eom}, the evolution equations for the internal degrees of freedom reduce to
\begin{equation} \label{eq:internal_iso_eom} \frac{d s_i}{dt} = \frac{1}{3}\frac{\partial \mathcal{H}^{\text{WP}}}{\partial \pi_{s,i}} \quad ; \quad \frac{d\pi_{s,i}}{dt} = -\frac{1}{3}\frac{\partial \mathcal{H}^{\text{WP}}}{\partial s_i}. \end{equation}
The factor of three is a direct result of transitioning from a three-by-three matrix to a single variable in Eq.~\eqref{eq:ani_to_iso}. This form is closely related to the isotropic Gaussian first used to model dense plasmas~\cite{Klakow1994a}, and subsequently studied by many authors.

\subsection{Additional potentials}

In WPMD models for dense plasma, additional effective potentials are typically included, see, e.g., Refs.~\cite{Svensson2023,Klakow1994a,Su2007}. First, so-called \emph{Pauli potentials} are additional two-body potentials motivated by a pairwise truncation of the antisymmetrization operator \cite{Klakow1994a,Grabowski2014}, partially addressing the limitation of the Hartree product in the restricted wavefunction. We employ Pauli potentials in this work, although directly antisymmetrized wavefunctions have previously been considered~\cite{Jakob2007}. Alternative schemes inspired by exchange-correlation energy approximations from DFT have also been investigated~\cite{Lavrinenko2021,Lavrinenko2019}. Secondly, a confining potential is used to enforce localized Gaussian wavepackets, which is standard in high temperature applications of WPMD. This is what we focus on in the present work. An additional potential energy operator is added to the electronic Hamiltonian~\cite{Svensson2024pub,Plummer2026}:
\begin{equation} \label{eq:electron_confining} \hat{V} = \frac{A}{2} \sum_i \hat{W}_i = \frac{A}{2} \sum_i (\hat{\boldsymbol{x}}_i - \langle \hat{\boldsymbol{x}}_i \rangle)^2 \end{equation}
where \(\hat{\boldsymbol{x}}_i\) denotes the position operator of the \(i\)-th electron and \(\langle\hat{\boldsymbol{x}}_i\rangle\) denotes its expectation value. We have also defined the width-squared operator \(\hat{W}_i\), which shall be used to characterize the extent of the electron wavefunction. As shall be demonstrated, adjusting \(A\) modulates the distribution of the width variables (\(\Sigma_i\) or \(s_i\)) in the model. This changes the effective Coulomb interactions and resulting structural properties~\cite{Plummer2026}. An alternative scheme based on a global confining potential to impose reflecting wall boundary conditions has also been studied in Refs~\cite{Lavrinenko2021, Lavrinenko2016}. We outline how the statistical framework can be extended to such potentials in Appendix~\ref{app:global_confining_potential}.

\section{Statistical theory for the width distribution}

In this section we develop a statistical theory for the internal degrees of freedom based on the non-interacting limit. Specifically, we assume single-particle contributions to the total energy are dominant and that the presence of width-dependent Coulomb and Pauli interactions has a small effect on the internal degrees of freedom. This assumption is correct for sufficiently weakly-coupled systems. More generally, its validity may be assessed by comparison to interacting molecular dynamics simulations, as performed in Sec.~\ref{sec:comparison_against_md}; results from these simulations are also used in Appendix~\ref{app:effective_coupling} to quantify the effective coupling $\Gamma^{\rm eff}$. Of course, at sufficiently strongly-coupled conditions, when $\Gamma^{\text{eff}} > 1$, the presence of interactions will perturb the width distribution away from the ideal case. Finally, we note that width distributions have been discussed for isotropic wavepackets in previous works, see e.g. Refs~\cite{Klakow1994b,Ma2019,Lavrinenko2016,Lavrinenko2019,Lavrinenko2021b}, but not directly predicted.

\subsection{Anisotropic case}
\label{sec:width_dist}
In an ideal gas of anisotropic wavepackets, the Hamiltonian is a sum over kinetic terms and confining potential terms:
\begin{equation} \label{eq:Ideal_WP_Hamiltonian} \mathcal{H}^{\text{WP}} = \sum_i \left(\mathcal{T}_i + \frac{A}{2}\text{Tr}\left\{\Sigma_i\right\}\right). \end{equation}
As discussed in the previous section, the final term represents the confining potential with empirical strength \(A\). This term arises from evaluating the expectation value of Eq.~\eqref{eq:electron_confining} for an anisotropic Gaussian wavefunction, given in Eq.~\eqref{eq:ani_gauss_wavefunction}. Without this term, the partition function diverges due to infinite wavepacket spread~\cite{Ebeling2006}. For anisotropic wavepackets the kinetic energy is given by~\cite{Svensson2023}
\begin{equation} \mathcal{T}^{\text{ani}} = \frac{\boldsymbol{p}^2}{2 m} + \frac{2}{m} \text{Tr}\left\{\Pi \Sigma \Pi\right\} + \frac{\hbar^2}{8m}\text{Tr}\left\{\Sigma^{-1}\right\} \end{equation}
where \(m\) is the electron mass and the particle index has been suppressed. In a non-interacting system with classical Hamiltonian equations, the many-body distribution function factorizes into a set of \(N\) single-particle distribution functions:
\begin{equation} \label{eq:ideal_single_particle_dist_func} f_1(\boldsymbol{p}, \Sigma, \Pi) = \frac{1}{Z_1} \exp\left[-\beta \left(\mathcal{T} + \frac{A}{2}\text{Tr}\left\{\Sigma\right\}\right)\right], \end{equation}
where \(\beta = 1/(k_B T)\) is the inverse temperature and \(Z_1\) is the single-particle partition function associated to the extended phase space of the model. Note that no closed-form expression is readily available for \(Z_1\). Introducing two-body interactions into Eq.~\eqref{eq:Ideal_WP_Hamiltonian} would couple the particle degrees of freedom, meaning that the Hamiltonian would no longer be separable into independent single-particle contributions. The full \(N\)-particle distribution function would therefore no longer factorize into single-particle terms, which is the central assumption underlying Eq.~\eqref{eq:ideal_single_particle_dist_func}. A complete treatment of the interacting system would require sampling this high-dimensional distribution, for instance using Monte Carlo methods.

To study the resulting distribution over widths, we proceed by integrating the single-particle distribution function over both the position and momentum-like degrees of freedom to find a marginal distribution in terms of the \(\Sigma\) matrix:
\begin{equation} \label{eq:changing_variables} f_{\Sigma}^{\text{ani}} (\Sigma) = \int d^3\boldsymbol{r} \int d^3\boldsymbol{p} \int d^6\Pi \, f_1(\boldsymbol{p}, \Sigma, \Pi). \end{equation}
The \( \Sigma \) matrix is symmetric, and therefore admits an eigendecomposition,
\(
\Sigma = \Lambda D \Lambda^\top
\),
where \( \Lambda \in \mathbb{R}^{3 \times 3} \) is an orthogonal matrix of eigenvectors (\( \Lambda^\top \Lambda = I \)) and \( D = \mathrm{diag}(\lambda_1, \lambda_2, \lambda_3) \) is a diagonal matrix of eigenvalues. Therefore, upon making the following change of variables,
\(
\Pi' = \Lambda^\top \Pi \Lambda,
\) the trace may be written as
\(
 \operatorname{Tr}(\Pi \Sigma \Pi) = \operatorname{Tr}(\Pi' D \Pi').
\)
Orthogonal transformations preserve volume and so the integration measure remains unchanged, i.e. 
\(
d^6\Pi = d^6\Pi',
\)
and the total distribution becomes
\begin{equation} f_\Sigma^{\text{ani}} (\Sigma) = \mathcal{Z} \exp\left[-\beta\left(\frac{\hbar^2}{8m}\text{Tr}\left\{\Sigma^{-1}\right\} + \frac{A}{2}\text{Tr}\left\{\Sigma\right\}\right)\right] I_1 , \end{equation}
where \(\mathcal{Z}\) is used henceforth as a normalization constant that is redefined as necessary to absorb prefactors. The last term is the following integral:
\begin{equation} I_1 = \int \exp\left[-\beta\frac{2}{m} \text{Tr}\left\{\Pi' D \Pi'\right\}\right] d^6\Pi'. \end{equation}
The matrix \( \Pi' \) is symmetric, so the trace can be expanded accordingly:
\begin{align*}
\operatorname{Tr}(\Pi' D \Pi') &= \sum_{p=1}^3 \lambda_p (\Pi_{pp}')^2 + \sum_{1 \le p < q \le 3} (\lambda_p + \lambda_q) (\Pi_{pq}')^2.
\end{align*}
Therefore the integral factorizes into six independent Gaussian integrals. Performing these integrals yields
\begin{widetext}
\begin{equation} f_{\Sigma}^{\text{ani}}(\Sigma) = \mathcal{Z} \left( \prod_{p=1}^{3} \frac{1}{\sqrt{\lambda_p}} \right) \left( \prod_{1 \le p < q \le 3} \frac{1}{\sqrt{\lambda_p + \lambda_q}} \right) \exp\left[-\beta\left(\frac{\hbar^2}{8m}\text{Tr}\left\{\Sigma^{-1}\right\} + \frac{A}{2}\text{Tr}\left\{\Sigma\right\}\right)\right]. \end{equation}
Since the shape matrix \(\Sigma\) is symmetric, it has six degrees of freedom. We wish to study the distribution of widths, which depend on the three eigenvalues of \(\Sigma\), denoted by \(\lambda_1, \lambda_2, \lambda_3\). To isolate these, we express the integral in terms of these three variables and integrate over the three rotation variables \(\mu_1, \mu_2, \mu_3\) which parameterize the possible orientation of the eigenbasis. It is possible to show that the Jacobian associated to this transformation is~\cite{Mehta2004}
\begin{equation} \mathcal{J}(\lambda_1, \lambda_2, \lambda_3, \mu_1, \mu_2, \mu_3) = \prod_{1 \le p < q \le 3} |\lambda_p - \lambda_q| g(\mu_1, \mu_2, \mu_3), \end{equation} where \(g\) is the distribution function of the \(\mu\) degrees of freedom. After integrating over the rotational degrees of freedom, the marginal distribution function over the eigenvalues is

\begin{equation} \label{eq:ani_lambda_dist} f_{\boldsymbol{\lambda}}^{\text{ani}}(\lambda_1, \lambda_2, \lambda_3) = \mathcal{Z} \left( \prod_{p=1}^{3} \frac{1}{\sqrt{\lambda_p}} \right) \left( \prod_{1 \le p < q \le 3} \frac{|\lambda_p - \lambda_q|}{\sqrt{\lambda_p + \lambda_q}} \right) \exp\left[ -\beta \left( \frac{\hbar^2}{8m} \sum_{p=1}^3 \frac{1}{\lambda_p} + \frac{A}{2} \sum_{p=1}^3 \lambda_p \right) \right]. \end{equation}
\end{widetext}
To characterize the resulting distribution we sample directly from the joint distribution of the eigenvalues \(\boldsymbol{\lambda} = (\lambda_1,\lambda_2,\lambda_3)\) defined in
Eq.~\eqref{eq:ani_lambda_dist}. From each set of sampled eigenvalues we calculate the widths 
\begin{equation} \sigma_p = \sqrt{\lambda_p} \quad \text{for} \quad p=1,2,3 \,, \end{equation} which provide a more natural physical interpretation. In order to characterize the statistics
of these widths we consider the distribution of an individual width, denoted $P^{\text{ani}}_\sigma$, obtained by pooling all $\sigma_p$ values across samples. Additionally, to measure the total wavepacket extent, we define the root-mean-squared (RMS) width as
\begin{equation} \label{eq:ani_averaged_width} \bar{\sigma} = \sqrt{\frac{1}{3}\langle\hat{W}\rangle} = \sqrt{\frac{1}{3}(\sigma_1^2 + \sigma_2^2 + \sigma_3^2)}, \end{equation}
where the first equality is the general definition in terms of the width-squared operator defined in Eq.~\eqref{eq:electron_confining}. Evaluating the state average for the specific case of anisotropic wavepackets yields the second equality. The marginal probability distribution of the RMS width $\bar{\sigma}$ is denoted $P^{\text{ani}}_{\bar{\sigma}}$. These two marginal distributions capture, respectively, the typical fluctuations of a single component and the collective behavior of the total width.

\subsection{Isotropic case}

The equivalent Hamiltonian for an ideal gas of isotropic wavepackets is
\begin{equation} \mathcal{H}^{\text{WP}} = \sum_i \left(\mathcal{T}_i + \frac{3A}{2}s_i\right), \end{equation}
which includes the state average of the confining potential using the isotropic wavepacket form given in Eq.~\eqref{eq:iso_gauss_wavefunction}. Here, \(s_i\) is the isotropic covariance parameter of the \(i\)th wavepacket. For isotropic wavepackets, the kinetic energy is
\begin{equation} \mathcal{T}^{\text{iso}} = \frac{\boldsymbol{p}^2}{2 m} + \frac{6}{m}\pi_s^2 s + \frac{3\hbar^2}{8m s}, \end{equation}
where the particle index is suppressed. Following the same procedure as the previous subsection and integrating over all momentum-like degrees of freedom gives the following distribution function over \(s\),
\begin{equation} \label{eq:iso_lambda_dist} f_{ s }^{\text{iso}}( s ) = \sqrt{\frac{3A\beta}{2\pi s }} \exp\left[ -\beta \left(\frac{3\hbar^2}{8 m s } + \frac{3}{2}A s - \frac{3}{2}\hbar\sqrt{\frac{A}{m}}\right)\right]. \end{equation}
Here the normalization factor has been explicitly calculated. The width metric employed for the anisotropic case can also be applied to the isotropic case. Specifically, we define the isotropic width variable 
\begin{equation} \label{eq:iso_averaged_width} {\sigma} = \sqrt{\frac{1}{3}\langle\hat{W}\rangle} = \sqrt{ s }, \end{equation}
{where the second equality is evaluated for the isotropic Gaussian, Eq.~\eqref{eq:iso_gauss_wavefunction}, in contrast to the anisotropic wavefunction used to evaluate the state average appearing in the second equality of Eq.~\eqref{eq:ani_averaged_width}. The width variable \(\sigma\) has the following marginal distribution function:
\begin{equation} f_\sigma^{\text{iso}}(\sigma) = 2\sigma f^{\text{iso}}_{ s }\left(\sigma^2\right), \end{equation}
which is the isotropic equivalent} {to the marginal probability of the averaged width \(P_{\bar\sigma}^{\text{ani}}\)}. It is also possible to find a closed-form expression for the mean width of the distribution,
\begin{equation}\label{eq:iso_mean}
\begin{split}
\langle \sigma \rangle^{\text{iso}} = & \int \sigma f_\sigma^{\text{iso}}(\sigma) d\sigma\\
= & \sqrt{\frac{3\hbar^2\beta}{2\pi m}}\exp\left[\frac{3}{2}\beta\hbar\sqrt{\frac{A}{m}}\right]K_1\left[\frac{3}{2}\hbar\beta\sqrt{\frac{A}{m}}\right]
\end{split}
\end{equation}
where \(\langle\cdot\rangle^{\text{iso}}\) denotes the statistical average with respect to the isotropic width distribution function and \(K_\alpha[\cdot]\) denotes the modified Bessel function of the second kind of order \(\alpha\). Additionally, the maximum of the distribution, calculated with
\begin{equation} {\frac{\partial f_{\sigma}^{\text{iso}}}{\partial \sigma} = 0}, \end{equation} gives the isotropic modal width:
\begin{equation} \label{eq:iso_mode} \text{mode}(\sigma) = \left(\frac{\hbar^2}{4mA}\right)^{1/4}, \end{equation}
which is equivalent to equating shape kinetic energy and confining potential terms, and has previously been used to ascribe an approximate electronic lengthscale to the confining strength in WPMD~\cite{Knaup1999,Plummer2026}. 

\section{Comparison against interacting molecular dynamics data} \label{sec:comparison_against_md}

\subsection{Numerical Details}
\subsubsection{Molecular dynamics}
We now compare the non-interacting theory developed in the previous section with fully interacting molecular-dynamics data for warm dense hydrogen. The full model being employed, including confining and Pauli potentials, corresponds to the fully ionized model specified in Ref.~\cite{Plummer2026}. The long-range Coulombic interactions are computed in reciprocal space according to the Ewald treatment in Ref.~\cite{Svensson2023}. A timestep of \(5\times10^{-5} \, \text{fs}\) and real-space cutoff for Pauli interactions of \(12 \, a_0\) ensures stable energy conservation. Results are presented for a Brueckner density parameter of \(r_s =2\) and a quantum degeneracy parameter of \(\theta= k_B T/ E_F = 1\), where \(E_F\) is the Fermi energy. Width distribution data are time-averaged over snapshots that encompass a \(25 \, \text{fs}\) period.

\subsubsection{Markov Chain Monte Carlo}
To compute the width distributions in the non-interacting statistical theory of anisotropic wavepackets, we sample \(\boldsymbol{\lambda} = (\lambda_1, \lambda_2, \lambda_3)\) from Eq.~\eqref{eq:ani_lambda_dist} using an MCMC sampling routine. Specifically, to generate samples from the target density we employ the \emph{random-walk Metropolis} algorithm~\cite{metropolis1953,hastings1970,robert2004}. At each step the algorithm proposes a move in parameter space and then accepts or rejects it according to the Metropolis rule. Concretely, starting from an initial state $\boldsymbol{\lambda}^{(0)}$, the algorithm iterates the following steps:
\begin{enumerate}
 \item Propose a new point
 $\boldsymbol{\lambda}^{\text{prop}}
 = \boldsymbol{\lambda}^{(t)} + \boldsymbol{\eta}$,
 where $\boldsymbol{\eta}\sim\mathcal{N}(\mathbf{0},\,\sigma_{\rm prop}^2 I)$.
 \item Compute the log of the unnormalized target density at the proposal
 and the current state, and form the acceptance probability
 \[\alpha = \min\!\left\{1,\,
 \exp\!\big[
 \log f^{\text{ani}}_{\boldsymbol{\lambda}}
 (\boldsymbol{\lambda}^{\text{prop}})
 - \log f^{\text{ani}}_{\boldsymbol{\lambda}}
 (\boldsymbol{\lambda}^{(t)})
 \big]\right\}.\]
 \item With probability $\alpha$, accept the proposal and set
 $\boldsymbol{\lambda}^{(t+1)}=\boldsymbol{\lambda}^{\text{prop}}$;
 otherwise, keep the current state
 $\boldsymbol{\lambda}^{(t+1)}=\boldsymbol{\lambda}^{(t)}$.
\end{enumerate}
Evaluating the density in log space prevents numerical underflow and
converts products into sums, which is more numerically stable.

The Gaussian proposal width \(\sigma_{\rm prop}\) is varied to yield a reasonable acceptance rate of around \(0.3\). Each independent chain consists of \(2\times 10^6\) iterations, with the first \(10^5\) discarded as burn-in and every 20th sample retained for analysis. For each parameter set, the retained samples from ten independent chains are combined to estimate the reported distributions. The retained states are transformed to \(\sigma_p=\sqrt{\lambda_p}\), from which we construct the single-particle distribution \(P^{\text{ani}}_{\sigma}\) and the averaged-width distribution \(P^{\text{ani}}_{\bar{\sigma}}\). Convergence was assessed using integrated autocorrelation times and Gelman--Rubin statistics across the ten chains. In the worst case across all sampled and derived quantities, the mean autocorrelation time was \(\tau \simeq 398\) MCMC steps~\cite{Gelman2013Ch11}, corresponding to a minimum effective sample size of \(N_{\rm eff}\simeq 4.8\times10^4\). The maximum Gelman--Rubin statistic was \(\hat{R}=1.000144\)~\cite{Gelman1992,Gelman2013Ch11}, close to the limiting value of unity expected for well-mixed chains. These diagnostics, combined with the good agreement shown in the following section, indicate robust convergence.

\subsection{Model comparison}
\begin{figure*}
 \includegraphics[width=\linewidth]{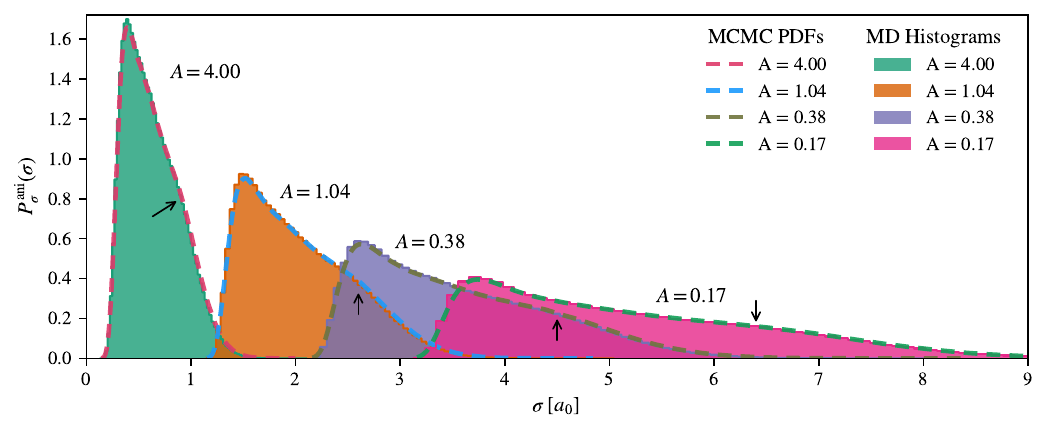}
 \caption{\label{fig:ani_width_dists} Marginal width probability distribution functions (PDFs) for an anisotropic wavepacket plasma at \(r_s=2\) and \(\theta=1\) for different confining potentials. Excluding the \(A=4.00 \, \text{Ha}/\text{a}_0^2\) case, each curve is shifted by $1 \, a_0$ relative to the curve to its left for clarity. The histograms correspond to molecular dynamics (MD) results in the microcanonical ensemble, while the dashed lines correspond to Markov Chain Monte Carlo (MCMC) sampling of Eq.~\eqref{eq:ani_lambda_dist}. Arrows indicate the shoulder feature discussed in the main text. The simulations are performed within a cubic box of side length \(L=52.5 \, a_0\). Hartree atomic units are used in the legend. }
\end{figure*}

\begin{figure}
 \includegraphics[width=\linewidth]{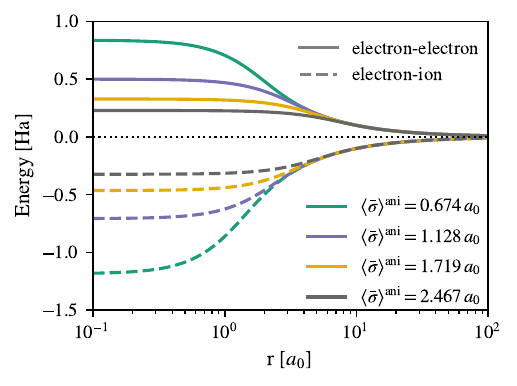}
 \caption{\label{fig:effective_interactions} Effective interparticle Coulomb interactions corresponding to the distributions plotted in Fig.~\ref{fig:ani_width_dists}. Calculation details for effective interactions are explained in Appendix~\ref{app:effective_coupling}}.
\end{figure}

\begin{figure}
 \includegraphics[width=\linewidth]{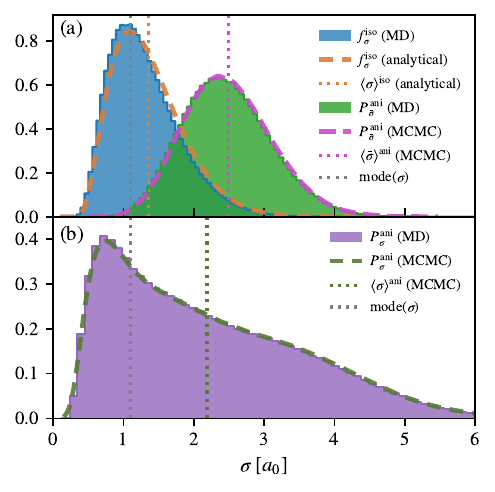}
 \caption{\label{fig:iso_and_ani_width_dist} Comparison between isotropic and anisotropic width distribution functions at \(r_s=2, \theta=1\) for a confining potential of \(A=0.17 \, \text{Ha}/\text{a}_0^2\). (a) Average width distribution functions for the isotropic and anisotropic models. (b) Distribution of individual width variables in the anisotropic model. The grey dashed line indicates the modal width of the isotropic model given by Eq.~\eqref{eq:iso_mode}. The mean values of each distribution are also plotted, with the isotropic case given by Eq.~\eqref{eq:iso_mean}, and the anisotropic case calculated numerically.}
\end{figure}

Fig.~\ref{fig:ani_width_dists} shows the sampled distributions compared to the interacting molecular dynamics simulations specified in the previous subsection for different values of the confining strength \(A\).
The agreement between the classical statistical theory and interacting width distributions is remarkable. Additionally, a shoulder feature is observed in the tail of each distribution, as indicated by the arrows in Fig.~\ref{fig:ani_width_dists}. This feature arises due to the eigenvalue difference term, \(\prod_{1 \leq p < q \leq 3} |\lambda_q - \lambda_p|\), present in the marginal distribution over eigenvalues, which appears in the second product of Eq.~\eqref{eq:ani_lambda_dist}. This \emph{eigenvalue repulsion} suppresses configurations with a similar spatial extent in each diagonalized spatial dimension, because fewer distinct covariance matrices \(\Sigma\) correspond to such points. This term directly originates from the rotational degrees of freedom in the model, which provide additional relaxation pathways. Further evidence for the link between the rotational degrees of freedom and the shoulder feature is provided in Appendix~\ref{app:diagonal_anisotropic_wavepacket}, where a wavepacket model with three independent widths but no rotational degrees of freedom is considered. The equivalence between the width distribution from the statistical theory and from the interacting system provides direct evidence that the effective interactions of the model are mediated by the classical statistics of the wavepacket Hamiltonian \(\mathcal{H}^{\text{WP}}\). To demonstrate this point we plot effective interparticle interactions in Fig.~\ref{fig:effective_interactions}. These interactions are computed from the Coulomb interactions of an isotropic Gaussian with a width given by the mean of the corresponding distributions in Fig.~\ref{fig:ani_width_dists}. For narrow wavepackets, the interparticle Coulomb interaction is greater at smaller separations given that the charge distribution is more localized. While for larger separations with respect to the wavepacket width, the pair potentials asymptote to the Coulomb potential.
At this condition, full ionization is expected, and the effect of interactions has little impact on the width PDFs. A comparison between the isotropic and anisotropic restricted wavefunctions is presented in Fig.~\ref{fig:iso_and_ani_width_dist}. 
Panel (a) shows the isotropic distribution \(f_\sigma(\sigma)\) computed with both MD and MCMC techniques, alongside the probability distribution for the average width of an anisotropic wavepacket \(P_{\bar{\sigma}}^{\text{ani}}\).
For an equivalent confining potential, the average extent of the anisotropic wavepacket is significantly larger. Additionally, the isotropic wavepacket distribution is slightly more localized than its corresponding ideal distribution, which can be attributed to the larger effective coupling for isotropic wavepackets, as quantified in Appendix~\ref{app:effective_coupling}. Panel (b) depicts the corresponding marginal distribution over individual width variables, which exhibits a prominent tail in the width distribution extending beyond \(6 \, a_0\) and a shoulder feature. Fig.~\ref{fig:iso_and_ani_width_dist} also demonstrates that the width of each anisotropic wavepacket is underestimated by the mode of the isotropic distribution. While emphasizing the role of the underlying classical statistics, this comparison elucidates the difference in the two restricted wavefunctions. The isotropic wavefunctions are considerably more localized for a given confining potential. This is likely explained by additional rotational degrees of freedom present in the anisotropic wavepacket ansatz that provide additional relaxation pathways in the dynamics, which manifest as eigenvalue repulsion. The stronger localization in the isotropic ansatz for a given confining strength means it will have stronger electron-electron and electron-ion interactions, explaining weaker correlations observed for the anisotropic ansatz in a previous comparison~\cite{Svensson2023}, albeit for a different form of the confining potential.

\section{Discussion}
\label{sec:discussion}

WPMD-based simulations can capture the concurrent time evolution of thousands of coupled ions and electrons. However, applications in plasma physics require the use of a confining potential to appropriately constrain the electron widths. In this work, we have presented a method by which the effect of the confining potential can be predicted \emph{a priori} for both isotropic and anisotropic variants of WPMD. From a practical perspective, this may allow confinement strength heuristics, such as those discussed in Refs~\cite{Ebeling2006,Knaup1999}, to be developed and then applied, without the need to explicitly scan the parameter using molecular dynamics simulations to find a suitable choice, as was required in Ref.~\cite{Plummer2025}. Additionally, we have investigated the distribution of electronic widths within a classical statistical theory. The comparison of two wavepacket models reveals that, even under equivalent confining potentials, they exhibit different mean widths and therefore different effective Coulomb interactions. This finding accounts for the structural differences previously reported between these models in the warm dense matter regime~\cite{Svensson2023}. The results presented here could also be extended to the case of a global external confining potential and therefore used to predict its impact on the equilibrium density profile~\cite{Lavrinenko2021}.

From a theoretical perspective, we have demonstrated that the distribution of variables in wavepacket models is closely connected to the statistical properties of the underlying classical Hamiltonian $\mathcal{H}^{\text{WP}}$. For the model to yield correct effective pairwise interactions--which are known from quantum statistical considerations to be Coulomb-like at large separations and finite at small separations in weakly-coupled, fully-ionized plasmas~\cite{Filinov2004}--it should relax to a non-divergent width distribution. As $A \rightarrow 0$ in either the anisotropic or isotropic case, the average width diverges and the distribution functions themselves, Eqs.~\eqref{eq:ani_lambda_dist} and \eqref{eq:iso_lambda_dist}, become unnormalizable. This observation establishes the confining potential as an essential ingredient for enforcing the correct behavior in weakly-coupled regimes. Similar models, such as those used for nuclear collisions, often address this feature by fixing the wavepacket width directly~\cite{Ono2004}. 

In fully ionized warm dense systems, as studied here, a significant fraction of the electrons are expected to be in spatially-extended states. However, a fully delocalized Gaussian wavepacket will effectively cease to interact with the system and has no mechanism to relocalize. Therefore, by regulating the model by choosing an appropriate confining potential, physically meaningful results can be obtained~\cite{Plummer2026,Svensson2024pub,Lavrinenko2021}.

\vspace{1mm}

\begin{acknowledgments}
 We are grateful for the use of computing resources provided by STFC Scientific Computing Department’s SCARF cluster. DP, PS, SMV and GG acknowledge support from AWE-NST UK via Oxford Centre for High Energy Density Science (OxCHEDs). PS acknowledges funding from the Oxford Physics Endowment for Graduates (OXPEG). S.M.V. acknowledges support from the UK EPSRC grant EP/W010097/1. The work of SMV and GG has received partial support from EPSRC and First Light Fusion under the AMPLIFI Prosperity partnership, grant no. EP/X025 373/1. The work by PS was partially supported by the European Research Council (ERC) under the European Union’s Horizon 2022 research and innovation programme (Grant agreement No. 101076233, ``PREXTREME”). Views and opinions expressed are however those of the authors only and do not necessarily reflect those of the European Union or the European Research Council Executive Agency. Neither the European Union nor the granting authority can be held responsible for them.
\end{acknowledgments}

\section*{Data Availability}

The data used to generate the figures in the main text are openly available~\cite{plummer2026StatisticsFigureData}.

\appendix

\section{Global confining potentials}
\label{app:global_confining_potential}

In the main text, a width-dependent confining potential that 
penalizes the spatial extent of each wavepacket is considered. Some authors have considered
a global external potential acting directly on the electronic coordinates~\cite{Lavrinenko2016,Lavrinenko2021}:
\begin{equation} \hat{V}_{\rm g} = \sum_i V_{\rm g}(\hat{\boldsymbol{x}}_i). \end{equation}
For a Gaussian wavepacket, this contributes to the wavepacket Hamiltonian
through the state average
\begin{equation} U_{\rm g}(\boldsymbol{r}_i,\Sigma_i) = \langle V_{\rm g}(\hat{\boldsymbol{x}}_i)\rangle = \int d^3\boldsymbol{x}\, \rho_i(\boldsymbol{x};\boldsymbol{r}_i,\Sigma_i) V_{\rm g}(\boldsymbol{x}), \end{equation}
where \(\rho_i\) is the single-particle probability density associated with the
Gaussian wavepacket. For example, in the case of anisotropic Gaussians,
\(\rho_i = |\langle\boldsymbol{x}|q_i\rangle|^2\). Continuing with the example of anisotropic wavepackets, the ideal single-particle distribution considered in Sec.~\ref{sec:width_dist} would then be replaced by
\begin{equation} f_1(\boldsymbol{r},\boldsymbol{p},\Sigma,\Pi) = \frac{1}{Z_1} \exp\left[ -\beta \left( \mathcal{T}^{\rm ani} + U_{\rm g}(\boldsymbol{r},\Sigma) \right) \right]. \end{equation}
The momentum degrees of freedom can still be integrated out, giving a
configurational distribution
\begin{equation} \label{eq:conf_PDF} f^{\mathrm{conf}}(\boldsymbol{r},\Sigma) = \int d^3\boldsymbol{p} \int d^6\Pi\, f_1(\boldsymbol{r},\boldsymbol{p},\Sigma,\Pi). \end{equation}
This object is the natural analogue of the configurational distribution used in
the main text. However, in contrast to the local width potential, a global
potential generally couples the expected position \(\boldsymbol{r}\) to the
shape variable \(\Sigma\) through \(U_{\rm g}(\boldsymbol{r},\Sigma)\).

The statistical distribution over widths is obtained by marginalizing
Eq.~\eqref{eq:conf_PDF} over the expected position coordinate. For example, the marginal
distribution over the covariance matrix is
\begin{equation} f_\Sigma(\Sigma) = \int d^3\boldsymbol{r}\, f^{\mathrm{conf}}(\boldsymbol{r},\Sigma). \end{equation}
Although the momentum integration can be performed,
a closed-form statistical distribution over widths is not guaranteed. This is
automatic for the local width potential considered in the main text, because the
expected position and width variables separate, but it is generally not true for a global
potential. For special choices of \(V_{\rm g}\), such as a global harmonic potential, the dependence on \(\boldsymbol{r}\) and \(\Sigma\) may separate. For more general global confinement schemes, including harmonic cubic box potentials, the coupling between the position and shape variables may prevent further
analytical progress. 

For models with external confining potentials, MCMC sampling of
\(f^{\mathrm{conf}}(\boldsymbol{r},\Sigma)\) provides a direct route to both the
statistical width distribution and the corresponding spatially varying ideal equilibrium
number density
\begin{equation} n(\boldsymbol{x}) = N_e \int d^3\boldsymbol{r} \int_{\Sigma>0} d^6\Sigma\, f^{\mathrm{conf}}(\boldsymbol{r},\Sigma) \rho_i(\boldsymbol{x};\boldsymbol{r},\Sigma), \end{equation}
where the \(\Sigma\) integral is restricted to positive-definite symmetric
matrices. This method may therefore be used to compare with published work on density distributions in global-confinement schemes~\cite{Lavrinenko2021}.

\section{Effective coupling parameter} \label{app:effective_coupling}

In this appendix we define an effective Coulomb coupling parameter that estimates the
role of interactions on the wavepacket width distribution. We first present the Coulomb interaction generated by Gaussian charge distributions and then
specialize the result to effective interactions for isotropic and anisotropic wavepackets. Consider two anisotropic Gaussian wavepackets with centres
\(\boldsymbol{r}_i\) and \(\boldsymbol{r}_j\), and shape matrices
\(\Sigma_i\) and \(\Sigma_j\). The probability density of the $i$th wavepacket is
\begin{equation} |\langle\boldsymbol{x}|q_i\rangle|^2 = \mathcal{N}^2 \exp\left[ -\frac{1}{2} (\boldsymbol{x}-\boldsymbol{r}_i)^T \Sigma_i^{-1} (\boldsymbol{x}-\boldsymbol{r}_i) \right], \end{equation}
with a similar expression for the $j$th wavepacket. The Coulomb interaction between wavepackets is
\begin{equation} \label{eq:coulomb_pair_potential} u_{ij}(\boldsymbol{R}_{ij}) = \frac{q_iq_j}{4\pi\epsilon_0} \int d^3x \int d^3y\, \frac{|\langle\boldsymbol{x}|q_i\rangle|^2 |\langle\boldsymbol{y}|q_j\rangle|^2} {|\boldsymbol{x}-\boldsymbol{y}|}, \end{equation}
where $\boldsymbol{R}_{ij} = \boldsymbol{r}_i - \boldsymbol{r}_j$ is the separation vector between the expected positions of the two wavepackets.

For isotropic wavepackets, \(\Sigma_i=\sigma_i^2 I\) and
Eq.~\eqref{eq:coulomb_pair_potential} reduces to the standard result
\begin{equation} u_{ij}^{\rm iso}(R) = \frac{q_iq_j}{4\pi\epsilon_0} \frac{\operatorname{erf}(\zeta_{ij}R)}{R} \quad ; \quad \zeta_{ij} = \frac{1}{\sqrt{2(\sigma_i^2+\sigma_j^2)}}. \end{equation}
For an electron--ion interaction with a point ion, one sets
\(\sigma_j=0\), giving
\begin{equation} \zeta_{iJ}^{\rm iso} = \frac{1}{\sqrt{2}\sigma_i}. \end{equation}

For an anisotropic electron interacting with a point-charge ion, the exact
interaction is obtained from Eq.~\eqref{eq:coulomb_pair_potential} with
\(|\langle\boldsymbol{y}|q_j\rangle|^2 \rightarrow \delta(\boldsymbol{y} - \boldsymbol{r}_j)\). While the anisotropic interaction does not generally reduce to a simple error-function form, numerical schemes have been developed based on a decomposition of the Coulomb interaction kernel~\cite{Svensson2023}. To obtain a scalar estimate that can be inserted into an
effective coupling parameter, the instantaneous width may be replaced by the statistical RMS width obtained from the width distribution, which for the electron--ion case yields
\begin{equation} \label{eq:effective_smoothing_electron_ion} \zeta^{\rm ani}_{i,J} = \frac{1}{\sqrt{2}\,\langle\bar{\sigma}\rangle^{\rm ani}} \quad , \quad \zeta^{\rm iso}_{i,J} = \frac{1}{\sqrt{2}\,\langle\sigma\rangle^{\rm iso}} \end{equation}
for anisotropic and isotropic wavepackets, respectively. Note that this replacement is not intended to reproduce the exact orientationally averaged anisotropic Coulomb interaction; it provides a scalar estimate based on the typical RMS width of the wavepacket. For an electron--electron estimate with two comparable wavepacket widths, \(\sigma_i\) and \(\sigma_j\) are both replaced by the corresponding statistical
width. This gives an additional factor of \(\sqrt{2}\) in the denominator of
Eq.~\eqref{eq:effective_smoothing_electron_ion}. To quantify the effective coupling, we use the electron--ion case because the interaction with a point ion has the shortest smoothing length and therefore gives the largest short-range Coulomb magnitude among interactions involving an electron wavepacket. We then define the following heuristic effective coupling parameter:
\begin{equation} \label{eq:effective_coupling} \Gamma^{\rm eff}(N_d,\zeta) = \frac{e^2}{4\pi\epsilon_0} \frac{\operatorname{erf}(\zeta a)}{a} \frac{1}{(N_d/3)k_B T}. \end{equation}
Here \(a = a_0 r_s\) is the characteristic interparticle separation. The quantity \(N_d\) is the
number of momentum degrees of freedom per particle, a factor that can be motivated from the equipartition theorem (see Appendix D in Ref.~\cite{Svensson2023}). For
point particles, \(N_d=3\) and \(\zeta\rightarrow\infty\), such that
Eq.~\eqref{eq:effective_coupling} reduces to the usual classical Coulomb
coupling parameter. For isotropic wavepackets $N_d = 4$, reflecting the inclusion of the additional momentum variable $\pi_s$. For anisotropic wavepackets $N_d = 9$, due to the presence of six degrees of freedom in the symmetric matrix $\Pi$. For Gaussian wavepackets, finite \(\zeta\) accounts for the regularization of
the Coulomb interaction at short range by the spatial extent of the electronic
charge density.

In Table~\ref{tab:effective_coupling} we present calculations of the effective coupling, finding that it lies below unity for the studied conditions. The isotropic wavepacket coupling is greater, which can be observed in Fig.~\ref{fig:iso_and_ani_width_dist} (a), where isotropic wavepackets are more localized than the noninteracting theory predicts.

\begin{table}[t] 
\caption{\label{tab:effective_coupling}
Computed effective Coulomb coupling parameters for isotropic and anisotropic
wavepackets. Width uncertainties are given in parentheses and denote one
standard deviation in the final quoted digits.
}
\begin{ruledtabular}
\begin{tabular}{ccccc}
\(A\) &
\(\langle\sigma\rangle^{\rm iso}\) &
\(\langle\bar{\sigma}\rangle^{\rm ani}\) &
\(\Gamma^{\rm eff}(4,\zeta^{\rm iso})\) &
\(\Gamma^{\rm eff}(9,\zeta^{\rm ani})\) \\
\hline
4.000 & 0.5183(4) & 0.6736(7) & 0.8166 & 0.3619 \\
1.041 & 0.7525(15) & 1.1282(17) & 0.8102 & 0.3353 \\
0.381 & 1.0126(27) & 1.7190(22) & 0.7772 & 0.2742\\
0.171 & 1.3051(36) & 2.4671(33) & 0.7142 & 0.2114 \\
\end{tabular}
\end{ruledtabular}
\end{table}

\section{Statistics of a diagonal-anisotropic wavepacket} \label{app:diagonal_anisotropic_wavepacket}

To isolate the role of independent width directions from the role of
rotational degrees of freedom, we consider an intermediate model with the following parameterization:
\begin{equation}
\begin{split}
\langle \boldsymbol{x} |h_i\rangle = \mathcal{N} \times \exp\bigg[& -\sum_{\alpha=1}^3 m_{i}^{(\alpha)} \left(x^{(\alpha)}-r_{i}^{(\alpha)}\right)^2\\
&+ \frac{i}{\hbar} \boldsymbol{p}_i^T(\boldsymbol{x}-\boldsymbol{r}_i) \bigg],
\end{split}
\end{equation}
where the $(\alpha)$ superscript denotes a vector component. As performed in Sec.~\ref{sec:Gaussian_wavepacket_model}, the complex width variable may be decomposed into real and imaginary parts,
\begin{equation} m_{i}^{(\alpha)} = \frac{1}{4s_{i}^{(\alpha)}} - \frac{i}{\hbar}\pi_{i}^{(\alpha)}. \end{equation}
This wavepacket can expand independently along the three Cartesian directions,
but its principal axes are fixed in the laboratory frame and therefore it
cannot rotate. We refer to this intermediate model as the
diagonal-anisotropic wavepacket. This model has not been considered in the main text, nor implemented in the MD code, because it unphysically constrains the wavepacket expansion along the three fixed axes and therefore imposes preferential expansion directions.

\begin{figure}
 \includegraphics[width=\linewidth]{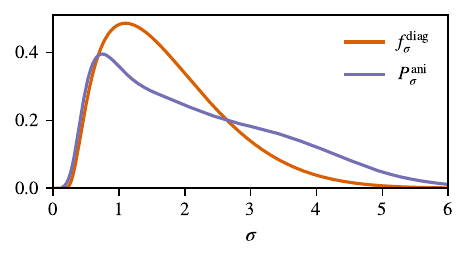}
 \caption{\label{fig:diagonal_versus_anisotropic_pdfs} Comparison between PDFs of diagonalized width variables for the diagonal-anisotropic wavepacket introduced in Appendix~\ref{app:diagonal_anisotropic_wavepacket} and the fully anisotropic wavepacket discussed in the main text. The conditions are the same as those in Fig.~\ref{fig:iso_and_ani_width_dist}.}
\end{figure}

Dropping the particle index, the RMS width of an arbitrary diagonal-anisotropic wavepacket is
\begin{equation} \bar{\sigma} = \sqrt{\frac{S}{3}} = \sqrt{\frac{1}{3}\left(s^{(1)} + s^{(2)} + s^{(3)}\right)}, \end{equation}
where the intermediate variable $S$ has been defined. Using the framework in Sec.~\ref{sec:width_dist}, we find the combined PDF of the three covariance variables $\boldsymbol{s} = (s^{(1)}, s^{(2)}, s^{(3)})$ as
\begin{equation}
\begin{split} \label{eq:f_S_product_form}
f_{\boldsymbol{s}}(s^{(1)}, s^{(2)}, s^{(3)}) = \prod_{\alpha=1}^3 f_s^{\mathrm{diag}}(s^{(\alpha)})
\end{split}
\end{equation}
where
\begin{equation}
\begin{split}
f_s^{\mathrm{diag}}(s^{(\alpha)}) = & \, \sqrt{\frac{A\beta}{2\pi s^{(\alpha)}}}\\
\times &\exp\left[-\beta \left(\frac{\hbar^2}{8ms^{(\alpha)}} + \frac{A}{2}s^{(\alpha)} - \frac{\hbar}{2}\sqrt{\frac{A}{m}} \right) \right].
\end{split}
\end{equation}
After changing variables from \(s\) to \(\sigma=\sqrt{s}\), one obtains the
single-direction width distribution \(f_\sigma^{\rm diag}(\sigma)\), which is
directly comparable with the marginal eigenvalue width distribution of the fully
anisotropic model. These distributions are plotted in Fig.~\ref{fig:diagonal_versus_anisotropic_pdfs} for the condition studied in Fig.~\ref{fig:iso_and_ani_width_dist}. The shoulder feature, present in the anisotropic model, no longer appears in the diagonal-anisotropic model. This feature, together with the extended distribution tail, can therefore be
attributed to the additional rotational degrees of freedom present in the full
anisotropic model.

To obtain a statistical distribution over RMS widths, the PDF of \(S\) is first considered. This may be obtained by marginalizing the joint density over
all configurations of \(\boldsymbol{s}\) with the same value of \(S\):
\begin{equation}
\begin{split}\label{eq:f_S_triple_integral}
f_S(S) = \int_0^\infty ds^{(1)} \int_0^\infty & ds^{(2)} \int_0^\infty ds^{(3)} f_{\boldsymbol{s}}(s^{(1)},s^{(2)}, s^{(3)})\\
& \times \delta\!\left(S-s^{(1)} - s^{(2)} - s^{(3)} \right).
\end{split}
\end{equation}
This PDF of $S$ may be written as a convolution. Explicitly, after substituting the product form~\eqref{eq:f_S_product_form}, the inner integral in Eq.~\eqref{eq:f_S_triple_integral} becomes
\begin{equation}
\begin{split}\label{eq:f_s_inner_integral}
&\int_0^{\infty}ds^{(3)} f_s^{\mathrm{diag}}(s^{(3)}) \delta\left(S-s^{(1)}-s^{(2)}-s^{(3)}\right)\\
&= f_s^{\mathrm{diag}}\!\left(S-s^{(1)}-s^{(2)}\right) \Theta\!\left(S-s^{(1)}-s^{(2)}\right),
\end{split}
\end{equation}
where $\Theta(\cdot)$ is the Heaviside step function, which arises from the
restriction that \(s^{(3)}\geq 0\). For \(S>0\), the domain of Eq.~\eqref{eq:f_s_inner_integral} is therefore
restricted to
\begin{equation} 0 \leq s^{(1)} \leq S, \qquad 0 \leq s^{(2)} \leq S - s^{(1)} . \end{equation} Using these results, Eq.~\eqref{eq:f_S_triple_integral} may be written
\begin{equation}
\begin{split}
f_S(S) = & \int_0^S ds^{(1)} \int_0^{S-s^{(1)}} ds^{(2)}\\
&\times f_s^{\mathrm{diag}}(s^{(1)})f_s^{\mathrm{diag}}(s^{(2)})f_s^{\mathrm{diag}}(S - s^{(1)}-s^{(2)}),
\end{split}
\end{equation}
where the limits arise due to the restriction to positive random variables. This form is equivalent to the following triple convolution~\cite[Sec.~17.12, entry~5]{gradshteyn2000}:
\begin{equation} \label{eq:triple_conv} f_S (S) = (f_s^{\mathrm{diag}} \star f_s^{\mathrm{diag}} \star f_s^{\mathrm{diag}}) (S) \end{equation}
where $\star$ denotes a convolution. Given that $s>0$ and we are considering a convolution, it is natural to use a Laplace transform approach to solve for $f_S$. Applying the transform to Eq.~\eqref{eq:triple_conv} yields
\begin{equation} \label{eq:laplace_conv_identity} \mathcal{L}\{f_S\}(\gamma) = \left[\mathcal{L}\{f_s^{\mathrm{diag}}\}(\gamma)\right]^3, \end{equation}
where
\begin{equation} \mathcal{L}\{f_s^{\mathrm{diag}}\}(\gamma) = \int_0^{\infty}\exp(-\gamma s) f_s^{\mathrm{diag}}(s) ds. \end{equation}
Using a known transformation~\cite[Sec.~3.471, entry~15]{gradshteyn2000}, we perform the Laplace transform of $f_s$ to find 
\begin{equation}
\begin{split}
&\mathcal{L}\{f_s^{\mathrm{diag}}\}(\gamma) =\\
&\sqrt{\frac{\beta A}{\beta A+2\gamma}} \exp\left[-\frac{\hbar\sqrt{\beta}}{\sqrt{2m}} \left(\sqrt{\gamma+\frac{\beta A}{2}} - \sqrt{\frac{\beta A}{2}} \right) \right].
\end{split}
\end{equation}
By cubing this result, it directly follows from Eq.~\eqref{eq:laplace_conv_identity} that
\begin{equation} \label{eq:laplace_f_S}
\begin{split}
\mathcal{L}\{f_S\}(\gamma) = & \left(\frac{\beta A}{2} \right)^{3/2} \exp\left[\frac{3\beta\hbar}{2}\sqrt{\frac{A}{m}} \right] \left(\gamma+\frac{\beta A}{2} \right)^{-3/2}\\
&\times \exp\left[-\frac{3\hbar\sqrt{\beta}}{\sqrt{2m}} \sqrt{\gamma+\frac{\beta A}{2}} \right].
\end{split}
\end{equation}

Applying the inverse Laplace transform then yields the required distribution. The required inverse transform identity can be derived from the standard Laplace transform~\cite[Sec.~17.13, entry~30]{gradshteyn2000}
\begin{equation} \label{eq:inverse_laplace_identity} \mathcal{L} \left\{ S^{-1/2} \exp\left(-\frac{a^2}{4S}\right) \right\}(\gamma) = \sqrt{\pi}\, \gamma^{-1/2} \exp\left(-a\sqrt{\gamma}\right), \end{equation}
where the Laplace transform is taken with respect to \(S\). After dividing both sides by \(\sqrt{\pi}\gamma\) and rearranging, we now apply an inverse Laplace transform and use the transform-of-an-integral rule~\cite[Sec.~17.12, entry~3]{gradshteyn2000}. This yields
\begin{equation} \label{eq:laplace_inverse_identity}
\begin{split}
\mathcal{L}^{-1} &\left\{\gamma^{-3/2} \exp\left(-a\sqrt{\gamma}\right) \right\}(S)\\
&=\frac{1}{\sqrt{\pi}} \int_0^S u^{-1/2} \exp\left(-\frac{a^2}{4u}\right) du\\
&= \frac{2\sqrt{S}}{\sqrt{\pi}} \exp\left(-\frac{a^2}{4S}\right) - a\, \operatorname{erfc} \left(\frac{a}{2\sqrt{S}} \right),
\end{split}
\end{equation}
where the final equality follows from an integration by parts.

\begin{figure}
 \includegraphics[width=\linewidth]{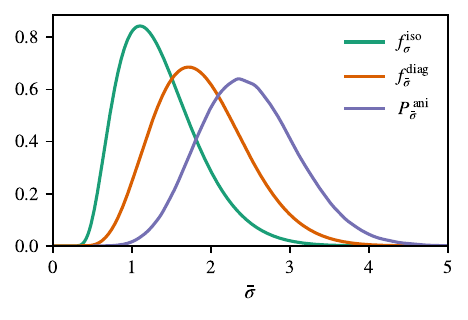}
 \caption{\label{fig:RMS_comparison_with_diagonal} Comparison between RMS width PDFs for isotropic, diagonal-anisotropic and anisotropic wavepackets. The conditions are the same as those in Fig.~\ref{fig:iso_and_ani_width_dist}.}
\end{figure}

We now use the derived identity, Eq.~\eqref{eq:laplace_inverse_identity}, to take the inverse Laplace transform of Eq.~\eqref{eq:laplace_f_S}. In this case,
\begin{equation} a=\frac{3\hbar\sqrt{\beta}}{\sqrt{2m}}, \end{equation}
and the argument of the transform is shifted according to
\[
\gamma \mapsto \gamma+\frac{\beta A}{2}.
\]
Using the shift theorem for Laplace transforms
\cite[Sec.~17.12, entry~4]{gradshteyn2000}, we therefore obtain
\begin{equation}
\begin{split}
& f_S(S) = \left(\frac{\beta A}{2} \right)^{3/2} \exp\left[\frac{3\beta\hbar}{2} \sqrt{\frac{A}{m}} - \frac{\beta A}{2}S \right]\\
&\times \left[\frac{2\sqrt{S}}{\sqrt{\pi}} \exp\left(-\frac{9\beta\hbar^2}{8mS} \right) - 6\sqrt{\frac{\beta\hbar^2}{8m}}\, \operatorname{erfc} \left(3\sqrt{\frac{\beta\hbar^2}{8mS}} \right) \right].
\end{split}
\end{equation}
Finally, to transform the PDF in \(S\) to the PDF in \(\bar{\sigma}\), we use \(
 S=3\bar{\sigma}^2,
\) and its derivative to write down the RMS-width distribution for diagonal-anisotropic wavepackets,
\begin{align}
& f_{\bar{\sigma}}^{\rm diag}(\bar{\sigma}) = 6\bar{\sigma} \left(\frac{\beta A}{2} \right)^{3/2} \exp\left[\frac{3\beta\hbar}{2}\sqrt{\frac{A}{m}} - \frac{3\beta A}{2}\bar{\sigma}^2 \right] \nonumber\\
& \, \times \left[\frac{2\sqrt{3}\bar{\sigma}}{\sqrt{\pi}} \exp\left(-\frac{3\beta\hbar^2}{8m\bar{\sigma}^2} \right) - 6\sqrt{\frac{\beta\hbar^2}{8m}}\, \operatorname{erfc} \left(\sqrt{\frac{3\beta\hbar^2}{8m\bar{\sigma}^2}} \right) \right].
\end{align}
This PDF for the RMS width is directly comparable with the isotropic and anisotropic equivalents, namely $f_\sigma^{\mathrm{iso}}$ and $P_{\bar{\sigma}}^{\mathrm{ani}}$. Comparing these distributions in Fig.~\ref{fig:RMS_comparison_with_diagonal}
shows that, as the number of degrees of freedom increases---from the isotropic model to the diagonal-anisotropic and finally to the anisotropic model---the RMS-width
distribution shifts toward larger widths, corresponding to a reduction in localization.

\clearpage

\bibliography{library, library_bound} 

\end{document}